\documentclass[pra,superscriptaddress,eqsecnum]{revtex4}
\usepackage{hyperref}
\usepackage{graphicx,color}

\begin{document}

\title{Validity of One-Dimensional QED for a System with Spatial Symmetry}
\author{Q.Z. Lv}
\affiliation{Intense Laser Physics Theory Unit and Department of Physics, Illinois State University, Normal, IL 61790-4560 USA}
\affiliation{State Key Laboratory for GeoMechanics and Deep Underground Engineering, China University of Mining and Technology, Beijing 100083, China}
\author{N.D. Christensen}
\affiliation{Department of Physics, Illinois State University, Normal, IL 61790-4560 USA}
\author{Q. Su}
\affiliation{Intense Laser Physics Theory Unit and Department of Physics, Illinois State University, Normal, IL 61790-4560 USA}
\author{R. Grobe}
\affiliation{Intense Laser Physics Theory Unit and Department of Physics, Illinois State University, Normal, IL 61790-4560 USA}
\date{\today}

\begin{abstract}
We examine the accuracy of an intrinsically one-dimensional quantum electrodynamics to predict accurately the forces and charges of a three-dimensional system that has a high degree of symmetry and therefore depends effectively only on a single coordinate.  As a test case we analyze two charged capacitor plates that are infinitely extended along two coordinate directions.  Using the lowest-order fine structure correction to the photon propagator we compute the vacuum's induced charge polarization density and show that the force between the charged plates is increased.  Although a one-dimensional theory cannot take the transverse character of the virtual (force-mediating) photons into account, nevertheless it predicts, in lowest order of the fine-structure constant, the Coulomb force law between the plates correctly.  However, the quantum correction to the classical result is slightly different between the 1d and 3d theories with the polarization charge density induced from the vacuum underestimated by the 1d approach. 
\end{abstract}

\maketitle

\section{Introduction}
	Spatially constrained models have been used rather successfully in basically all areas of physics.  Due to their restricted degrees of freedom they usually are computationally more feasible than their 3d counterparts and often provide a conceptually easier access to exploring complicated dynamics.  A good example is the strong-field ionization physics of atoms and molecules, where spatially constrained models have provided us with a wealth of qualitative information about the details of the multi-electron ionization paths \cite{DiMauro:2000}, the associated generation of higher harmonics \cite{LHuillier:1992} and various stabilization phenomena \cite{Su:1990}.  Here the spatial dependence of the atomic 1d binding potentials was chosen to mimick the energy spectra of their real 3d counterparts \cite{Su:2005}.  On a more fundamental level, the so-called 1d quantum field theories have been used widely \cite{Glimm:1968,Jaffe:1968,Glimm:1970} to overcome problems usually associated with mass and charge renormalization, to tackle conceptual difficulties and also to test the feasibility of new numerical approaches \cite{Wagner:2013coa}.  For example, 1d field theories have been used rather recently to study the electron-positron pair creation process induced by an supercritical external field with full space-time resolution \cite{Cheng:2010,Hebenstreit:2013baa,Hebenstreit:2013qxa,Steinacher:2014bda}.  In lowest-order perturbation theory, one-dimensional quantum electrodynamics predicts some peculiar features such as a position-independent Coulomb force between two 1d charges.  However, this theory is obviously not able to take accurately the transverse character of the photons into account.  In fact, a magnetic field cannot play any role in a one-dimensional world.  The most prominent example of a spatially reduced field theory is possibly the Schwinger model of QED \cite{Schwinger:1962tp}, where as an additional approximation it was assumed that the fermionic mass vanishes.  As a result, this model becomes an interacting quantum field theory that can be studied non-perturbatively.

	To the best of our knowledge, we are not aware of any quantitative study that compares the predictions of 1d quantum electrodynamics directly with its 3d counterpart for exactly the same physical system.  In order to do so we use a test system of charges that has a spatial symmetry in the $x_1-x_2$ plane such that any macroscopic observable depends at most on the $(x_3\equiv)z$-direction.  We can then apply the 3d theory to this highly symmetric system and compare the corresponding fields, forces and induced charges with the predictions of a theory, that is intrinsically one-dimensional.  The latter is defined and obtained from the fundamental form of the 3d theory by neglecting any derivative with respect to $x_1$ and $x_2$ from the very beginning.

	This paper is organized as follows:  In Section~\ref{sec:classical}, we introduce the model system of two plane parallel capacitor plates and use classical electrodynamics to compute the Coulomb forces and charges from the 1- and 3-dimensional approaches.  In Section~\ref{sec:3d QED}, we use 3d QED to derive the analytical expressions for the corrections of these forces due to the occurrence of the vacuum's polarization charges close to the plates.  In Section~\ref{sec:1d QED}, we derive the one-dimensional QED and apply it to the two-plate system.  For a better structure, we have shifted the mathematically more complete derivations to the appendices.  In Section~\ref{sec:discussion}, we compare directly the predictions of the 1d and 3d approaches.  Section~\ref{sec:summary} summarizes this work and motivates several future studies.

\section{\label{sec:classical}The Classical Two-Plate System}
\begin{figure}
\begin{center}
\includegraphics[scale=0.7]{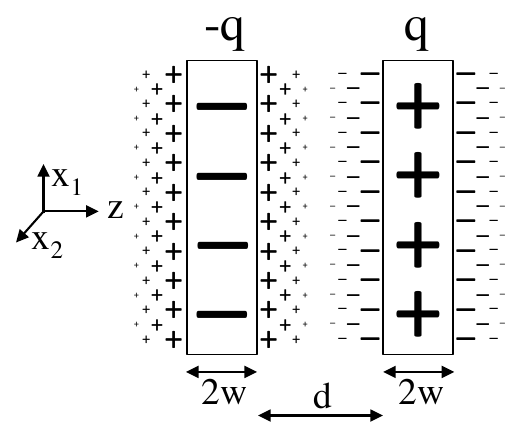}
\end{center}
\caption{\label{fig:plates}Sketch of the infinitely extended plane-parallel capacitor plates with width 2w each and edge-to-edge spacing d.  The parameter q is the (two-dimensional) charge density (charge per unit area in the $x_1-x_2$ direction).  We also sketch the vacuum's induced polarization charges.}
\end{figure}

	To have a concrete quantitative example for our analysis, we examine two plane parallel plates that are separated by a distance denoted by $d$.  We assume that each plate has a width of $2w$ and is infinitely extended along the $x_1$- and $x_2$-directions as seen in Figure~\ref{fig:plates}.  For simplicity, we also assume that the three-dimensional (unperturbed) charge density (measured in $C/m^3$) is constant on each plate, described by the density $\rho(\vec{r}) = q/(2w) [-U_w(z+d/2+w)+U_w(z-d/2-w)]$, where we denote by $U_w(z)$ the rectangular unit step function, defined as $U_w(z)\equiv 1$ for $|z|<w$ and $U_w(z)\equiv 0$ for $w<|z|$.  Here we denote with $q\equiv Q_0/L^2$ the (positive) two-dimensional charge density.  Both the total charge $Q_0$ and the area $L^2$ of each plate are infinite, but the density $q$ (measured in $C/m^2$) is finite.  In our analysis, we assume that the total charge $Q_0$ on the right plate, is chosen to be independent of the width $2w$ along the z-direction, $Q_0 = \int d x_1 d x_2\int_{d/2}^{d/2+2w}dz\ q/(2w) = q L^2$.

	Before we apply quantum field theory, let us first neglect any effect of the vacuum's polarizability and, as an introduction, use the classical Maxwell equations to derive the force per unit area between the plates.  Since in a classical treatment neither the force nor the energy associated with the two-plate system depend on $w$, we can assume in this section the limit $w=0$.  The scalar potential $\phi(\vec{r})$ associated with the (positively charged) right plate located at $z=d/2$, can be obtained from the stationary Maxwell equation  $-\nabla^2\phi(\vec{r}) = 4\pi k_e \rho(\vec{r})$, where the associated charge density is $\rho(\vec{r}) = q \delta(z-d/2)$.  We abbreviate Coulomb's constant as $k_e\equiv 1/(4\pi\varepsilon_0)$, which is related to the vacuum's permittivity $\varepsilon_0$.  We obtain $\phi(\vec{r}) = - 2\pi k_e q |z-d/2|$, where we choose the convention that the potential vanishes at the boundary, $\phi(x_1,x_2,z=d/2) = 0$.  The associated electric field follows as $E(\vec{r}) = -\nabla \phi(\vec{r}) = 2\pi k_e q (z-d/2)/|z-d/2|$ and is constant outside the plate. 

	We note that the combined electric field of both plates (at $z=\pm d/2$) vanishes outside the plates,  $|z| > d/2$, and takes the constant (negative) value $- 4\pi k_e q$ between the plates.  The fact that the electric field vanishes outsides the plates seems to suggest naively that it should be impossible to induce any polarization charges outside the plates.  Interestingly, we will argue in the discussion below that this conjecture is incorrect when the effect of the quantum vacuum is accurately taken into account.

	The (negative) work it takes to move the left plate (of finite but large area $L^2$) with charge density $-q$ from location $z=-\infty$ to $z=-d/2$ is therefore $V(d) = \int d^3r (-q) \delta(z+d/2) \phi(r) = 2 \pi k_e q^2 d L^2$, leading to a finite attractive force per area of magnitude $F(d)/L^2  = \partial_d V(d)/L^2 = 2\pi k_e q^2$ between the two plates.  Note that this classical Coulomb force $F(d)$ does not depend on the separation $d$ between the plates. 

	Let us now introduce an intrinsically 1d description from a classical perspective.  In the more general context of QED this description will be derived slightly differently and more rigorously in Section~\ref{sec:1d QED}.  For simplicity, we assume here that the corresponding 1d Maxwell equation for the 1d potential takes the same functional form as its 3d counterpart, except that we have removed the derivatives with respect to the two extraneous coordinates $x_1$ and $x_2$, leading to $- \partial_z^2 \phi_{1d}(z) = 4\pi k_e q \delta(z-d/2)$.  We note that in this particular (preliminary) approach to a 1d theory, the potential $\phi_{1d}(z)$ (with the subscript 1d rather than the superscript 1d of Section~\ref{sec:1d QED} and Appendix~\ref{app:1d QED}) and $\phi(r)$ would have the same units of $J/C$.  
As expected, we obtain $\phi_{1d}(z) = -2\pi k_e q |z-d/2|$ as in the 3d theory.  However, the corresponding 1d interaction energy between the "one-dimensional point charges" $V_{1d}(d)$ defined here as $V_{1d}(d)\equiv \int dz (-q) \delta(z+d/2) \phi_{1d}(z) = 2\pi k_e q^2 d$ has different units ($J/m^2$) than $V(d)$ ($J$) due to the lack of the factor $L^2$ that we have to include in all 1d energies and forces to become comparable to the real 3d system.  

	While the classical finding that the 1d- and 3d- approach predict the same Coulomb force law $F(d)/L^2  = - 2\pi k_e q^2$ remains valid also in a more rigorous quantum description (to the lowest order in the fine-structure constant), we will show below that in order to derive a consistent 1d quantum theory from the 3d theory, the potentials $\phi_{1d}(z)$ and $\phi(r)$ are required to take different units.

\section{\label{sec:3d QED}3D QED for the Plate System}
Let us first compute the effect of the vacuum's polarization density and the electric field for a single (positively charged) plate centered at $z=0$.  In this context we note that even the sign of the charges induced from the vacuum seems to be controversial in the literature \cite{note1,Peskin:1995ev,Greiner:1985}.  The lowest-order correction term to the Feynman photon propagator can be obtained from the diagram shown in Figure~\ref{fig:modified photon prop} in Appendix~\ref{app:quantum correction to potential}.


Although the natural unit system is intuitive and typically used in quantum field theory, we keep here the SI units as we believe that they better illustrate the differences between a 1d and a 3d theory.  This is especially true, as the derivations contain Green's functions whose effective singular source term is usually chosen to have different units than the real physical sources for the fields.
 
	Using dimensional regularization and also charge renormalization, we show in Appendices~\ref{app:Feynman Rules} and \ref{app:quantum correction to potential} that the modified photon propagator $D'_F$ due to the vacuum polarization for the 3d system takes the form \cite{Weinberg:1995mt,Ryder:1985wq,Srednicki:2007qs,Schwartz:2013pla}:
\begin{equation}
D'_{F\mu\nu}(k)  = D_{F\mu\nu}(k)\left(1 + \alpha \left[P(k^2) - P(0)\right]\right)
\end{equation}
where $\alpha\equiv k_e e^2/(\hbar c)$, $c$ is the speed of light and $\hbar$ is Planck's constant.  The remaining (unitless) integral is defined as
\begin{equation}
P(k^2) - P(0) = \frac{2}{\pi} \int_0^1 d\beta \beta (1-\beta) \mbox{ln}\left[1-\beta(1-\beta)k^2\lambda_C^2\right]
\end{equation}
where $\lambda_C\equiv\hbar/(m c) = 3.86\times10^{-13}$m is the electron's reduced Compton wavelength and $m$ is the mass of the electron.  A rather lengthy computation, which we review in Appendix~\ref{app:quantum correction to potential}, leads to the Uehling potential \cite{Uehling:1935uj} for a general three-dimensional charge configuration given by the density $\rho(\vec{r}^{\ \prime})$.  It takes the form
\begin{equation}
\phi(\vec{r})  =  k_e\int d^3r' \frac{\rho(\vec{r}^{\ \prime})}{|\vec{r}-\vec{r}^{\ \prime}|} 
\left[1 + \frac{\alpha}{3\pi}\int_1^\infty d\tau f(\tau)e^{-2\tau |\vec{r}-\vec{r}^{\ \prime}|/\lambda_C}\right]
\label{eq:secIII:phi(r)}
\end{equation}
where we introduced the abbreviation $f(\tau) \equiv \left(2/\tau^2+1/\tau^4\right)\sqrt{\tau^2-1}$.  The Compton wavelength characterizes the relevant length scale of the polarization charge cloud.

	We can apply Eq.~(\ref{eq:secIII:phi(r)}) to our (positively charged) plate (of width $2w$ and centered around $z=0$) for which the (unperturbed) charge density is given by $\rho(\vec{r}^{\ \prime}) = q/(2w) U_w(z')$.  The first (classical) term in Eq.~(\ref{eq:secIII:phi(r)}), which is the corresponding unperturbed Coulomb potential $\phi^{(0)}$ for the plate, can be evaluated in cylindrical coordinates as $\phi^{(0)}(\vec{r}) = 2\pi k_e \int dr' dz' r'\rho(z')/\sqrt{r'^2+(z-z')^2}$ and leads to
\begin{equation}
	    \phi^{(0)}(\vec{r}) = - 2\pi k_e q \frac{z^2+w^2}{2w} U_w(z) -2\pi k_e q |z| \left[1-U_w(z)\right]
\label{eq:secIII:phi0}
\end{equation}
The second term of Eq.~(\ref{eq:secIII:phi(r)}) (due to the vacuum polarization) leads to 
\begin{eqnarray}
\phi^{(1)}(\vec{r}) &=&
-2\pi k_e\lambda_C^2 \frac{\alpha}{12\pi} \frac{q}{2w} \int_1^\infty d\tau \frac{f(\tau)}{\tau^2}
\left[ -2 + e^{2\tau(z-w)/\lambda_C} + e^{-2\tau(z+w)/\lambda_C} \right] U_w(z)\nonumber\\
&&+ 2\pi k_e \lambda_C^2 \frac{\alpha}{12\pi} \frac{q}{2w} \int_1^\infty d\tau \frac{f(\tau)}{\tau^2}
e^{-2\tau|z|/\lambda_C} \left[e^{2\tau w/\lambda_C} - e^{-2\tau w/\lambda_C}\right] \left[1-U_w(z)\right]
\label{eq:secIII:phi1}
\end{eqnarray}
This result can be effectively interpreted as being due to a polarization charge density due to the vacuum polarization correction, which can be determined from the classical Maxwell equation as $\rho_{pol}(\vec{r}) =-(4\pi k_e)^{-1}\nabla^2\phi^{(1)}(\vec{r})$.  We obtain:
\begin{eqnarray}
\rho_{pol}(\vec{r})   &=& \ \frac{\alpha}{6\pi} \frac{q}{2w} \int_1^\infty d\tau f(\tau)
\left[e^{2\tau(z-w)/\lambda_C} + e^{-2\tau(z+w)/\lambda_C} \right] U_w(z)\nonumber\\
&&    - \frac{\alpha}{6\pi} \frac{q}{2w} \int_1^\infty d\tau f(\tau) e^{-2\tau|z|/\lambda_C} 
\left[e^{2\tau w/\lambda_C} - e^{-2\tau w/\lambda_C}\right] \left[1-U_w(z)\right]
\label{eq:secIII:rho pol}
\end{eqnarray}
 One can easily see that the total amount of the induced charge vanishes $\int d^3r \rho_{pol}(\vec{r}) = -(4\pi k_e)^{-1}\int d^3r \nabla^2\phi^{(1)}(\vec{r}) = 0$, if the potential falls off rapidly enough as the distance from the plate goes to $\infty$, which it does.  We also note that the vacuum also seems to modify the charge distribution inside the plate itself ($-w<z<w$) in addition to the induced negative charge cloud around the plate.  This arises naturally in the quantum theory as the corrections to the Coulomb potential fall off exponentially so that the full potential (including quantum corrections) still falls off asymptotically as $\phi(r) = e k_e/r$ for $r\to\infty$ for a given physical charge $e$.
In order for the potential of the modified charge distribution to have this property, the actual charge located at $r=0$ has to be larger than $e$ such that the sum of the central charge and the negative charges around it amount to $e$.

As a side issue, we note that this static screening situation may be different than a hypothetical case where we would start with a given central charge and only afterwards turn on (artifically) the coupling to the vacuum.  In this time-dependent polarization scenario it is possible that the central charge actually remains the same and only negative polarization charges are induced around it \cite{Lv:tbp}.  In this particular dynamical polarization scenario, the total charge would decrease.

	In case of the plate, we can compute the extra amount of charge due to the polarization on the plate itself from Eq.~(\ref{eq:secIII:rho pol}) 
\begin{eqnarray}
Q_{pol} &\equiv& \int d x_1 d x_2 \int_{-w}^w d z \rho_{pol}(r) \nonumber\\
&=&  \frac{\alpha}{12\pi}  q L^2 \frac{\lambda_C}{w} \int_1^\infty d\tau \frac{f(\tau)}{\tau}
\left[1 - e^{-4\tau w/\lambda_C}\right] .
\label{eq:secIII:Qpol}
\end{eqnarray}
In contrast to the unperturbed total charge $Q_0 \equiv q L^2$, it is interesting to see that the induced positive charge on the plate itself $Q_{pol}$ increases monotonically as we decrease the width $2w$.  In other words, in the limit of an infinitesimally narrow plate ($w\to0$), we have $Q_{pol}\to\infty$ even for a finite $L$, which makes a quantitative analysis more difficult.  The same behavior is also observed for a 3d point charge, where the induced polarization charge "on top of" the original (unperturbed) finite positive charge is also infinite.

	In the opposite (and more intuitive) limit where $w\gg \lambda_C$ the argument of the exponential is sufficiently large and negative so that we can use $\int_1^\infty d\tau f(\tau)/\tau = 9\pi/16$ and approximate the total induced (positive) charge on the plate itself as 
\begin{equation}
Q_{pol} = \frac{3\alpha}{64}  q L^2  \frac{\lambda_C}{w} \quad         \mbox{for}\quad w\gg \lambda_C
\end{equation}

	Next we return to the two-plate system.  We assume that the first (positively charged) plate is now centered at $z = d/2$, while the left (negatively charged) plate is centered at $z=-d/2$, as sketched in Figure~\ref{fig:plates}.  
We take the limit $w\to0$ for simplicity of the final analytical expressions.
Similarly to the classical case, the total energy lost to move the left plate from minus infinity to location $z=-d/2$ can be evaluated as $V(d) = \int d^3r \rho(r) \left[\phi^{(0)}(r) + \phi^{(1)}(r)\right]$, where $\phi(r)$ and $\rho(r)$ are given by Eqs.~(\ref{eq:secIII:phi0}), (\ref{eq:secIII:phi1}) and (\ref{eq:secIII:rho pol}).  In order to be consistent with the order $O(\alpha)$ of our computation of $\rho_{pol}(r)$, we neglect the interaction between the induced charges.  We obtain
\begin{eqnarray}
  \frac{V(d)}{L^2}  &=&  2\pi k_e q^2 d 
- 2\pi k_e q^2  \frac{5\lambda_C \alpha}{6\pi} \int_1^\infty d\tau \frac{f(\tau)}{\tau} e^{-2\tau d/\lambda_C}
\end{eqnarray}
The final attractive force per area of $F(d)/L^2 = - \partial_dV(d)/L^2$ between the two plates can now be computed as:
\begin{equation}
  \frac{F(d)}{L^2}  = - 2\pi k_e q^2 
\left[1 + \frac{5\alpha}{3\pi} \int_1^\infty d\tau f(\tau) e^{-2\tau d/\lambda_C}\right]
\label{eq:secIII:F/L2}
\end{equation}

	We discuss the distance-dependence of this modified Coulomb force law for the two plates in Section~\ref{sec:discussion} in more detail. 

\section{\label{sec:1d QED}1D QED for the plate system}
\begin{figure}
\begin{center}
\includegraphics[scale=0.45]{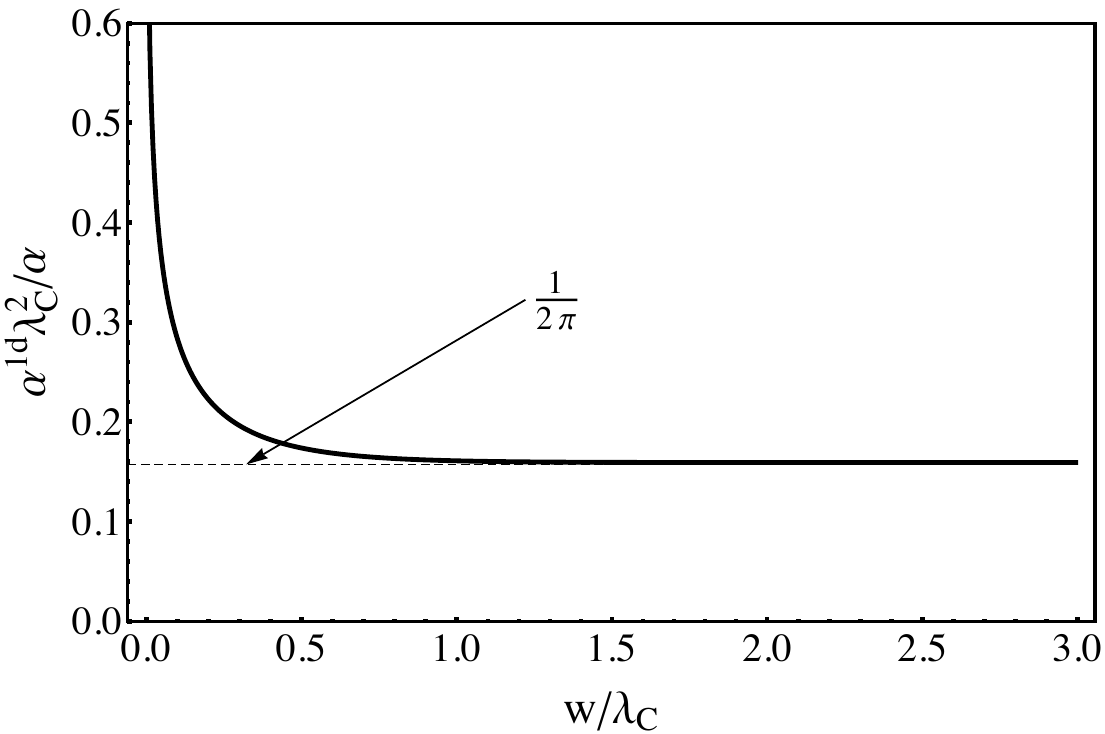}
\end{center}
\caption{\label{fig:alpha 1d / alpha}The numerical value of the fine-structure constant $\alpha^{(1d)}$ (in units of $\alpha/\lambda_C^2$) for one-dimensional QED as a function of the width $2w$ of the plate, based on the requirement that the total induced polarization charge from the vacuum on the plate should be identical for the 1d and 3d QED.}
\end{figure}

	In Eq.~(\ref{eq:appC:phi1d}) of Appendix~\ref{app:1d QED} we have derived the one-dimensional potential $\phi^{(1d)}(z)$ [now with a superscript 1d to distinguish it from the $\phi_{1d}(z)$ of Sec.~\ref{sec:classical}] for a general 1d charge distribution $\rho(z')$ as 
\begin{equation}
\phi^{(1d)}(z) = 2\pi k_e \int dz'\rho(z') \left[-|z-z'| +
\alpha^{(1d)}\lambda_C^3\int_1^\infty d\tau f^{(1d)}(\tau)e^{-2\tau|z-z'|/\lambda_C}\right] 
\label{eq:secIV:phi1d}
\end{equation}
where we have introduced the abbreviation $f^{(1d)}(\tau) \equiv 1/\left(\tau^5\sqrt{\tau^2-1}\right)$ and $\alpha^{(1d)}$ for the effective fine-structure constant of 1d.  The (unitless) function $f^{(1d)}(\tau)$ plays a similar role as $f(\tau)$ in the 3d case.  In order to make the quantitative connection with a three-dimensional plate (of width $2w$ and charge density $q$) located at $z=0$, we assume that the 1d density used in Eq.~(\ref{eq:secIV:phi1d}) is given by $\rho^{(1d)}(z') = q/(2w) U_w(z') L$.  Here the additional factor $L$ is required by the 3d to 1d translation rules between the charges as outlined in Appendix~\ref{app:1d QED}.  As a result, $\rho(r')$ (for 3d) and $\rho^{(1d)}(z')$ (for 1d) have different units.  While (except for the factor of $L$) the first (classical) part of the potential $\phi^{(1d,0)}(z)$ in Eq.~(\ref{eq:secIV:phi1d}) is identical to that of the infinite plate given in Eq.~(\ref{eq:secIII:phi0}), i.e., $\phi^{(1d,0)}(z) = L \phi^{(0)}(r)$, the second (vacuum polarization) part is different and becomes
\begin{eqnarray}
\phi^{(1d,1)}(z) &=&
- \pi k_e \alpha^{(1d)} \lambda_C^4 \frac{q}{2w} L \int_1^\infty d\tau \frac{f^{(1d)}(\tau)}{\tau}
\left[-2 + e^{2 \tau (z-w)/\lambda_C} + e^{-2\tau(w+z)/\lambda_C} \right] U_w(z)\nonumber\\
&&-\pi k_e \alpha^{(1d)} \lambda_C^4 \frac{q}{2w} L \int_1^\infty d\tau \frac{f^{(1d)}(\tau)}{\tau}
  \left[e^{-2\tau w/\lambda_C} -  e^{2\tau w/\lambda_C} \right] e^{-2\tau |z|/\lambda_C} \left[1-U_w(z)\right]
\end{eqnarray}
The corresponding contribution to the charge density due to the vacuum polarization can be calculated again from the classical (Maxwell equation) as $\rho^{(1d)}_{pol}(z) =-(4\pi k_e)^{-1}\partial_z^2 \phi^{(1d,1)}(z)$.  We obtain:
\begin{eqnarray}
\rho^{(1d)}_{pol}(z)  &=&
\alpha^{(1d)}\lambda_C^2 \frac{q}{2w} L \int_1^\infty d\tau \tau f^{(1d)}(\tau)
    \left[ e^{2\tau (z-w)/\lambda_C} + e^{-2\tau (w+z)/\lambda_C} \right] U_w(z)\nonumber\\
&&- \alpha^{(1d)} \lambda_C^2 \frac{q}{2w} L  \int_1^\infty d\tau \tau f^{(1d)}(\tau)
   \left[e^{2\tau w/\lambda_C} - e^{- 2\tau w/\lambda_C} \right] e^{-2\tau |z|/\lambda_C} \left[1-U_w(z)\right]
\label{eq:secIV:rho pol}
\end{eqnarray}
	Before we can compare these predictions with the result of the 3d theory, we have to determine first the unknown value of the 1d fine-structure constant $\alpha^{(1d)}\equiv k_e(e^{(1d)})^2/(\hbar c)$.  We propose here to determine this constant (and therefore the fundamental charge $e^{(1d)}$ of one-dimensional QED) by requiring that the total induced 1d charge [$Q_{pol}^{(1d)}\equiv \int_{-w}^w dz \rho_{pol}^{(1d)}(z)$] on the plate (when multiplied by $L$) has to match the corresponding total induced charge obtained from the 3d theory [$Q_{pol}\equiv \int_{-w}^w dz \rho_{pol}(z) L^2$].  If we then compare Eq.~(\ref{eq:secIII:Qpol}) with the spatial integral over Eq.~(\ref{eq:secIV:rho pol}) this equality simplifies to
\begin{equation}
\frac{\alpha}{6\pi} \int_1^\infty d\tau\frac{f(\tau)}{\tau}  \left(1- e^{-4\tau w/\lambda_C}\right)  
=  \alpha^{(1d)} \lambda_C^2 \int_1^\infty d\tau f^{(1d)}(\tau)\left(1 - e^{-4\tau w/\lambda_C}\right)
\label{eq:secIV:alpha comparison}
\end{equation}
Using this required equality, we can obtain the value of the 1d fine-structure constant $\alpha^{(1d)}$ as a function of $\alpha$, the Compton wavelength $\lambda_C$ and the plate width $w$.  We have graphed in Figure~\ref{fig:alpha 1d / alpha} the parameter $\alpha^{(1d)}$ according to Eq. (\ref{eq:secIV:alpha comparison}) as a function of $w$ and find that it approaches a constant value if the plate's width $w$ is larger than $\lambda_C$, which is a natural limit for a physical plate.

In the limit of small ($\lambda_C/w$) we can neglect the two exponentials in Eq.~(\ref{eq:secIV:alpha comparison}) and use $\int_1^\infty d\tau f(\tau)/\tau = 9\pi/16$ and similarly $\int_1^\infty d\tau f^{(1d)}(\tau) = 3\pi/16$.  As a result, we obtain 
\begin{equation}
  \alpha^{(1d)}  =  \frac{\alpha}{2\pi\lambda_C^2}
\label{eq:secIV:alpha comparison 2}
\end{equation}
We note that our required match in Eq.~(\ref{eq:secIV:alpha comparison 2}) determines also the value of the charge of the fundamental particle of the 1d QED world as $e^{(1d)} = e/(\lambda_C\sqrt{2\pi}) = 1.66\times10^{-7}$C/m.

	Knowing the value for the constant $\alpha^{(1d)}$, we can finally compare the predictions of the 1d theory quantitatively with the results from the actual 3d approach.  For example, similarly as in Sec.~\ref{sec:3d QED}, we can now compute the total energy per area between the two plates (again in the limit $w\to0$) from this 1d QED theory as:
\begin{equation}
\frac{V^{(1d)}(d)}{L^2}  = 2\pi k_e q^2 d 
- 3\pi k_e q^2 \alpha^{(1d)} \lambda_C^3 \int_1^\infty d\tau f^{(1d)}(\tau) e^{-2\tau d/\lambda_C}
\end{equation}
and correspondingly the total force per unit area between the two plates can be computed leading to: 
\begin{equation}
  \frac{F^{(1d)}(d)}{L^2}  
= - 2\pi k_e q^2  
\left[ 1 + 3 \alpha^{(1d)}\lambda_C^2 \int_1^\infty d\tau f^{(1d)}(\tau) \tau e^{-2\tau d/\lambda_C}\right]
\label{eq:secIV:F/L2}
\end{equation}

\section{\label{sec:discussion}Discussion}
	In this section we will compare directly the differences between the two approaches.  The charge renormalization and the subsequent fixing of the corresponding charge of the positronic elementary particle in 1d have guaranteed that the total induced charge per area on each plate is identical in both approaches for $w\gg\lambda_C$.
  
\begin{figure}
\begin{center}
\includegraphics[scale=0.4]{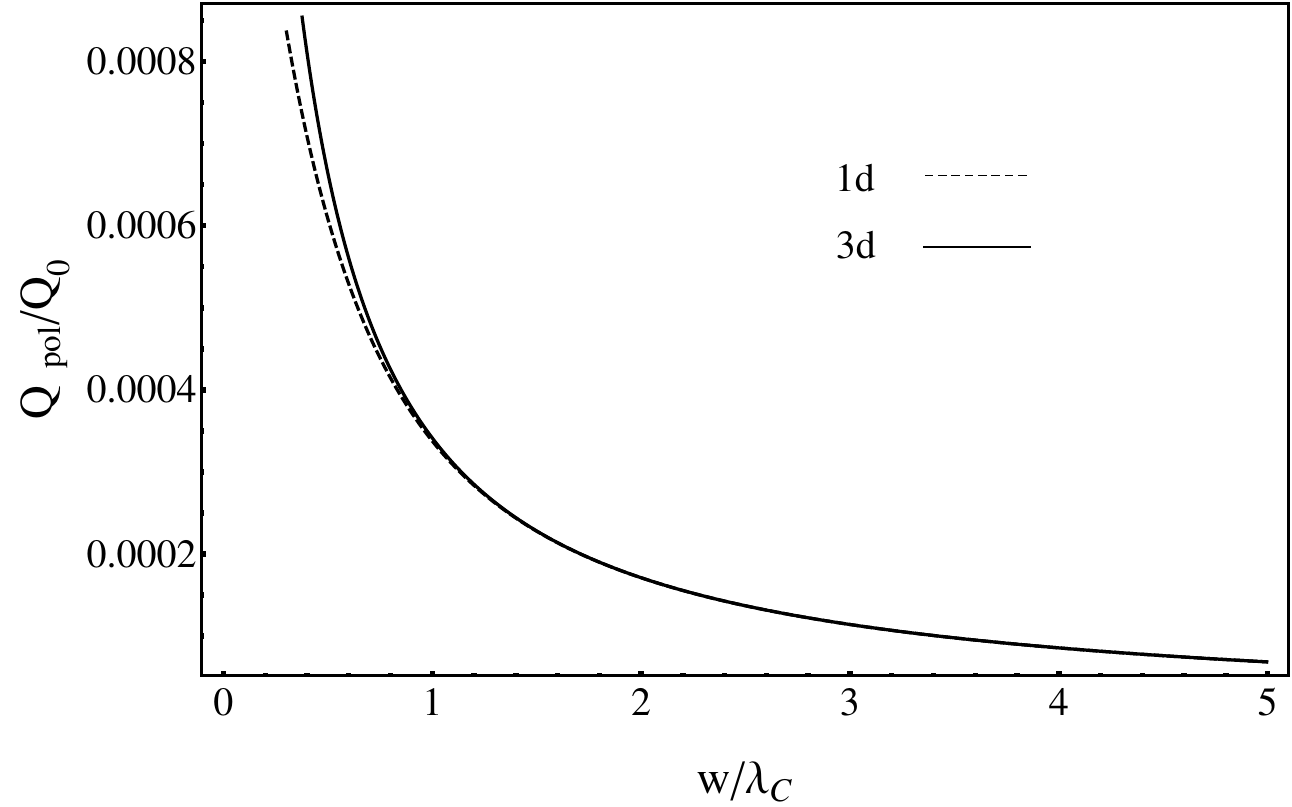}
\end{center}
\caption{\label{fig:Qpol/Q0}The ratio between the induced polarized charge $Q_{pol}$ and the initial charge $Q_0$ as a function of the width of the plate $w$.  The solid line is obtained from the 3d QED approach and the dashed line is the result of the 1d calculation.  The unperturbed density of the plate was $q = 1.602\times10^{-4} C/m^2$.}
\end{figure}

	First, let us discuss the ratio of the induced charge $Q_{pol}$ and the unperturbed charge $Q_0$ as a function of the width of the plate $w$ in Figure~\ref{fig:Qpol/Q0}.  The qualitative behavior obtained from the 1d and 3d theories match rather well if the plate is not too narrow.  Both approaches predict the same decrease of the polarization charge with increasing width $w$.  The difference between the two curves becomes apparent only for very narrow plates $w/\lambda_C < 1$, where the 3d theory predicts an infinite $Q_{pol}$ as $w/\lambda_C\to0$ while the 1d polarization charge remains finite.  We note that the 1d theory underestimates the polarization charge for narrow plates.
 
\begin{figure}
\begin{center}
\includegraphics[scale=0.4]{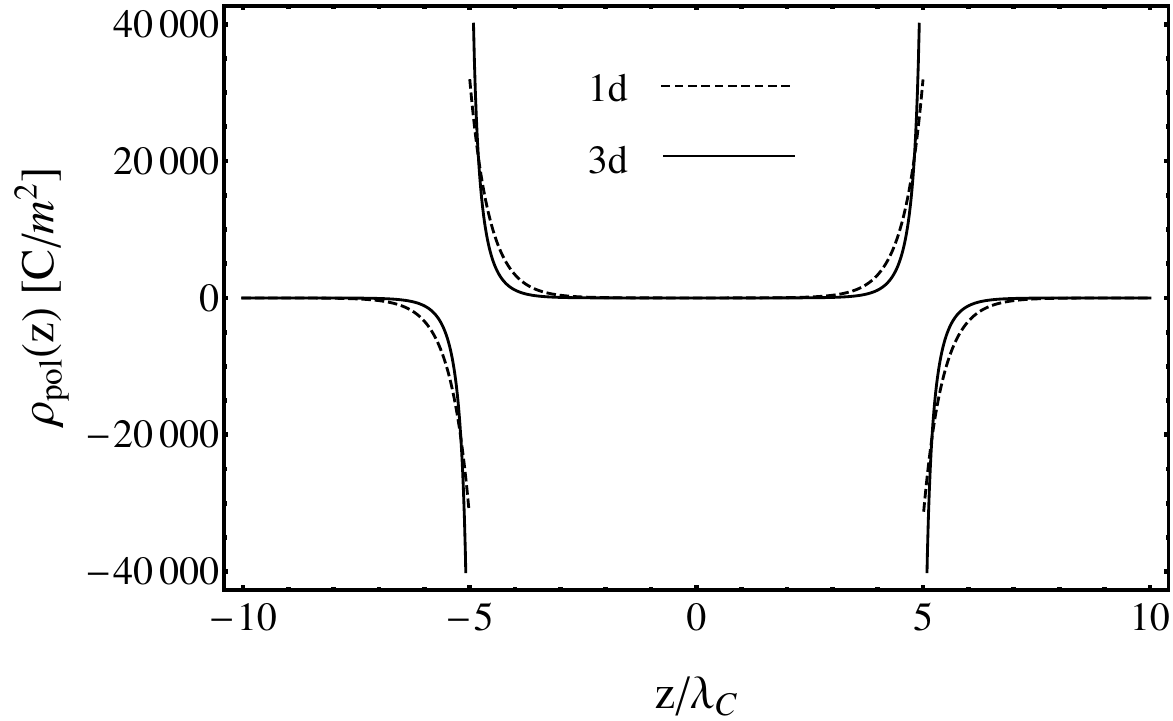}
\end{center}
\caption{\label{fig:rho pol}The polarization charge density $\rho_{pol}(z)$ measured in $C/m^2$ around the positive charged plate according to the three-dimensional (solid line) and the one-dimensional theory (dashed line) as a function of the position $z$.  The unperturbed density of the plate was $q= 1.602\times10^{-4} C/m^2$ and the plate has a width of $2w= 1.213\times10^{-11}$ m. }
\end{figure}

	In Figure~\ref{fig:rho pol}, we analyze the spatial dependence of the polarization charge for a plate centered at $z=0$ and width $2w=10\lambda_C$. 
 
	While the total induced charge $Q_{pol}$ is finite (unless $w/\lambda_C\to0$) for both theories, we see that the 3d theory predicts an infinite discontinuity between the induced charge densities $\rho_{pol}(z)$ at the edges of the plate $z=\pm w$.  In the 1d approach, however, this discontinuity is finite.  Both approaches predict the same spatial scale proportional to $\lambda_C$ at which the polarization charge density falls off as the distance from each edge increases. 

\begin{figure}
\begin{center}
\includegraphics[scale=0.45]{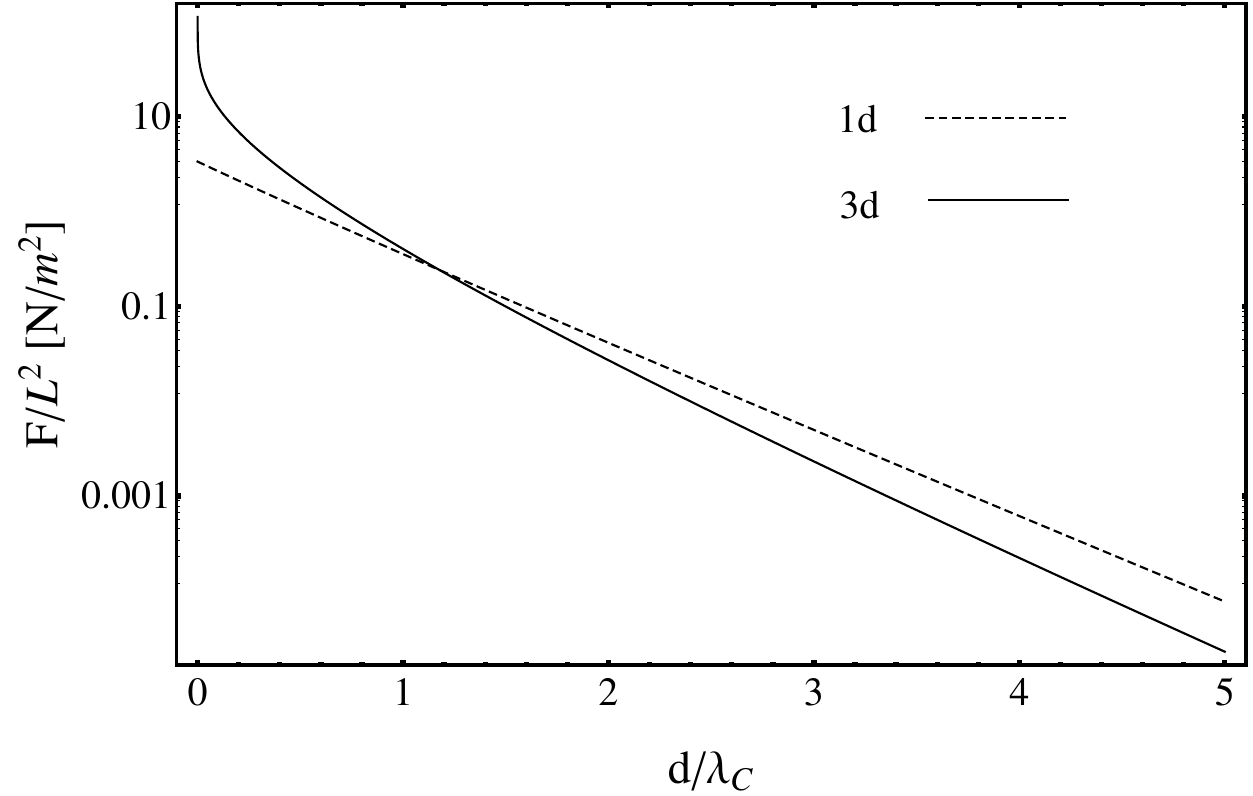}
\end{center}
\caption{\label{fig:F/A}The correction to the force per unit area (in $N/m^2$) between the two oppositely charged capacitor plates according to the three-dimensional (solid line) and the one-dimensional theory (dashed line) as a function of the distance $d$ between them.  The unperturbed density of each plate is $q= \pm 1.6\times10^{-4}  C/m^2$ and each plate has a width of $2w= 1.213\times10^{-11}m$.  For comparison, the classical force is $1450 N/m^2$.}
\end{figure}

	In Figure~\ref{fig:F/A}, we return to the two-plate geometry and discuss the modification to the usual classical Coulomb force law, derived in Section~\ref{sec:classical} as $F/L^2  = - 2\pi k_e q^2$.  We have compared numerically the modification of the force between the plates according to Eqs.~(\ref{eq:secIII:F/L2}) and (\ref{eq:secIV:F/L2}) as a function of the spacing $d$ for the two theories.  For the (arbitrary) parameters used in the figure ($q =1.6\times10^{-4} C/m^2$) the spacing-independent force (per unit area) between the plates would be $1450 N/m^2$ in the absence of any polarization.
 
	We show in Figure~\ref{fig:F/A} only the correction to the attractive classical force $F^{(1d)}(d)/L^2 = - 2\pi k_e q^2$.  We see from Eqs.~(\ref{eq:secIV:F/L2}) and (\ref{eq:secIII:F/L2}) that this correction increases the amount of the attractive force.  Had we chosen two equally charged plates we would have also observed that the vacuum would increase the repulsive force.

	The effect of the polarization charges on the force is obviously largest for small plate spacing d when the polarization charge clouds of both plates overlap the most.  For example, the 3d theory of Eq.~(\ref{eq:secIII:F/L2}) predicts an infinite force correction for $d=0$ [as $\int_1^\infty d\tau f(\tau)\to\infty$], while the 1d theory predicts a finite value according to Eq.~(\ref{eq:secIV:F/L2}), $6\pi k_e q^2 \alpha^{(1d)} \lambda_C^2\int_1^\infty d\tau f^{(1d)}(\tau)\tau$.  However, it should be kept in mind that the Uehling potential and our calculations are only strictly valid for distances on order of or greater than $\lambda_C$.

	For large plate spacings the 1d theory overestimates the correction due to the force, while due to its finite limit for $d=0$, it underestimates it for $d\to0$.  For a possible experimental test it is important to point out, the ratio of the correction to the force and the classical force is independent of the charge density $q$ and amounts to $5\alpha/(3\pi) \int_1^\infty d\tau f(\tau)\mbox{exp}(-2\tau d/\lambda_C)$.  This ratio decreases from infinity (for $d=0$) to $0.14$\% for $d/\lambda_C=0.5$.

	The nearly straight lines in the figure for $d/\lambda_C>1$ also suggest that it is possible to approximate the correction to the force by simple exponential functions of the distance.  If we approximate $\int_1^\infty d\tau f(\tau) \mbox{exp}(-2\tau d/\lambda_C)$ as $0.64\ \mbox{exp}(- 2.421 d/\lambda_C)$ and $\int_1^\infty d\tau f^{(1d)}(\tau)\tau \mbox{exp}(-2\tau d/\lambda_C)$ as $0.6\ \mbox{exp}(-2.127 d/\lambda_C)$, we would obtain simpler expressions for the correction force due to the polarization 
\begin{eqnarray}
  \frac{F_{pol}(d)}{L^2}     &=&  - 2\pi k_e q^2 \frac{5\alpha}{3\pi} 0.64 e^{- 2.421 d/\lambda_C}\\
  \frac{F^{(1d)}_{pol}(d)}{L^2} &=&  - 2\pi k_e q^2 \frac{3\alpha}{2\pi}  0.60 e^{-2.127 d/\lambda_C}
\end{eqnarray}

\section{\label{sec:summary}Summary and Open Questions}
	The purpose of this work was three-fold, to rigorously derive a 1d theory of QED, to test it for a real system with high spatial symmetry and to examine if it would be experimentally feasible to use a macroscopic system such as a two plane parallel capacitor system to measure the quantum correction to the classical Coulomb force between the two plates due to the vacuum polarization.  Even though any 1d theory is not able to describe the intrinsically transverse nature of the force mediating photons, it is surprising that (in the absence of any vaccum polarization) the 1d and 3d approaches predict an identical force.  This agreement is related mathematically to the fact that the unperturbed 3d Feyman photon propagator $D_{F\mu\nu}(x)$ when integrated over the extraneous coordinates $x_1$ and $x_2$ is identical to the corresponding 1d propagator of the 1d theory.  That is, these propagators have the property that
\begin{eqnarray}
  \int dx_1 dx_2 D_{F\mu\nu}(x) 
&=&  D^{(1d)}_{F\mu\nu}(x)\\
  \int dx_1 dx_2 S_F(x) 
&=&  S^{(1d)}_F(x)
\end{eqnarray}
where we have introduced our notation and derived the form of the photon and electron propagators $D_{F\mu\nu}(x)$ and $S_F(x)$ in Appendix~\ref{app:Feynman Rules}.

	The first-order (in the fine-structure constant) correction to the photon propagator is related to the product  $D_{F\mu\rho} \mbox{Tr}[V^\rho S_F V^\sigma S_F] D_{F\sigma\nu}$.  The difference between this and the analogous product based on the corresponding 1d-propagators is partly due to the fact that the integral ($\int  dx_1dx_2$) over a product is different than the product of individual integrals. Nevertheless, our final comparison suggests that the difference in our particular case of two sufficiently wide plates is not too large.  The natural question then might be if it is possible to construct a 1d theory that compensates for these differences and is able to agree completely at the next-to-leading order level with the 3d theory for systems of high $(x_1,x_2)$ symmetry.  Perhaps it is possible to analyze the magnitude and effect of the terms that are thrown away in the 1d Lagrangian and find a way to add compensating terms to the 1d Lagrangian without ruining its 1d computational benefits.

	There also remains the conceptual question of the induced polarization density in the region outside the two plates of opposite charge.  While classical electrodynamics predicts an identically vanishing E-field outside the plates, our calculations nevertheless show a non-vanishing induced charge.  Its presence might bring into question whether the illustrative picture of an unperturbed electric field that induces charges from the vacuum is really an appropriate framework.  

Moreover, it is also not fully understood whether the vacuum correction really causes the physical polarization of the vacuum by inducing physical charges that could be, in principle, measured.  To emphasize this point, we note that we obtained the vacuum charge density by taking the quantum calculation of the potential at next-to-leading order and forcing it back into a purely classical theory, which only corresponds to leading order.  It is not clear to us that this effective picture actually corresponds with Nature.  On the other hand, in principle, one could obtain the same result by computing the vacuum expectation value of the charge density operator directly \cite{Wichmann:1956zz,Fullerton:1976fu,Soff:1988zz,Neghabian:1983,Indelicato:2014mra}.  This seems to lend support to the idea of an actual physical vacuum charge density.  However, more work is required to resolve this puzzle.



Furthermore, while the whole concept of the vacuum's polarizabilty is based on a view point that is based on the existence of virtual charges, we point out that there are also proposed formalisms based on dressed particle states \cite{Greenberg:1900zz,Walter:1970,Stefanovich:2001gf,Stefanovich:2005ai,Stefanovich:2005hz} that do not require any virtual or bare particles.  It would be very interesting to examine in future work how the vacuum's polarizability would manifest itself in such alternative theoretical frameworks.  

	The effect of the force between two conducting plates due to the mode structure of the electromagnetic vacuum has been predicted \cite{Casimir:1948dh,Brown:1974pt,Harris:2000zz,Decca:2005,Gusso:2004hi,deMan:2009zz} and experimentally confirmed first \cite{Bressi:2002fr} for the Casimir effect.  In contrast to our case here, the two plates experiencing the Casimir force are uncharged and the force can be rather significant and even cause unwanted challenges in the manufacturing of small scale nano electromechanical materials \cite{note2,Lamoreaux:1996wh,Mohideen:1998iz,Chan:2001zzb}.  The effect of charges (such as highly charged ions) on the structure of the vacuum has been observed only in spectroscopic measurements asssociated with the energy shifts of certain energy levels.   In our macroscopic system of two charged parrallel plates, on the other hand, the amount of the quantum corrected force due to the polarization charges can be controlled by the amount of charge placed on the plates.  Nevertheless, the correction to the usual (plate-spacing independent) Coulomb force is only significant for extremely short distances on the order of the electron's Compton wavelength, making a direct experimental measurement of the force difficult for present technology.  

\section{Acknowledgements}
We thank S. Hassani and R.F. Martin, Jr. for helpful discussions.  We also enjoyed numerous discussions with A. Di Piazza, F. Fillion-Gourdeau, K. Hatsagortsyan, F. Hebenstreit, C. M\"uller and C. Schubert at KITP in Santa Barbara.  Q.Z.L. would like to thank ILP and the ISU physics department for the nice hospitality during his visit.  This work has been supported by the NSF, by the National Basic Research (973) Program of China (\#2013CBA01504), and the NSFC (\#11374360).  It also used the Extreme Science and Engineering Discovery Environment (XSEDE), which is supported by the NSF grant OCI-1053575.

\appendix

\section{\label{app:Feynman Rules}The Feynman Rules of QED in SI Units}
	As the final goals of our Feynman-rules based pertubative QED calculation are physical observables with SI units, we must first translate those rules from the so-called natural units of textbooks into SI units.  Although this process is straight forward, there are subtleties involving factors of $\hbar$ and $c$ that must be carefully worked out for correct diagrammatic rules.  In particular, in contrast to the situation of natural units, where the coefficient of the interaction terms of the Lagrangian are the same in its spatial and momentum form (equal to the bare charge $e_b$), in SI units, on the other hand, the Fourier transform of the Lagrangian from a spatial form to a momentum form introduces factors of $\hbar$ and $c$.  Moreover, the subtlety arises because the vertices and propagators are not observable, and unlike in the case of observables where the units of the observable determine the appropriate factors of $\hbar$ and $c$, we have no such guidance here.  For this reason, we give a complete derivation of the Feynman rules of QED in SI units in this appendix.  We furthermore note that, to the best of our knowledge, the SI form of the Feynman rules for QED can not be found in the literature.  Our derivation is based on the functional derivative, which is an equivalent alternative to the Wick contractions method found in some textbooks.  Below, we will show that the vertex is given by 

\begin{minipage}{3in}
\begin{center}
\includegraphics[scale=0.4]{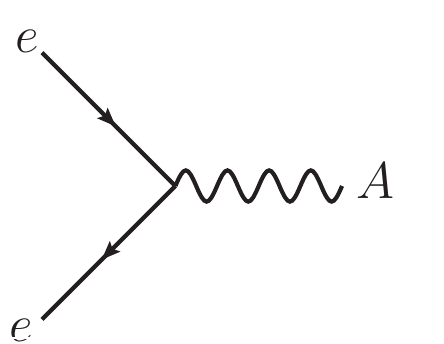}
\end{center}
\end{minipage}
\hfill
\begin{minipage}{3.8in} 
\begin{equation}
	V_\mu = -i \frac{e_b}{\hbar} \gamma_\mu
\end{equation}
\end{minipage}
and the propagators are given, in the Feynman gauge, by\\
\begin{minipage}{3in}
\begin{center}
\includegraphics[scale=0.6]{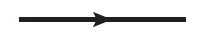}
\end{center}
\end{minipage}
\hfill
\begin{minipage}{3.8in}
\begin{equation}
S_F(k) = \frac{i}{\gamma^\mu k_\mu - m c/\hbar}
\end{equation}
\end{minipage}
\begin{minipage}{3in}
\begin{center}
\includegraphics[scale=0.6]{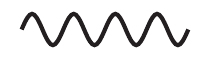}
\end{center}
\end{minipage}
\hfill
\begin{minipage}{3.8in}
\begin{equation}
D_{F\mu\nu}(k) = -i\frac{4\pi k_e \hbar}{c}\frac{\eta_{\mu\nu}}{k^2} .
\end{equation}
\end{minipage}
\\

From these expressions, we see that the units of the fermion propagator $S_F(k)$ is $m$, while the photon propagator $D_{F\mu\nu}(k)$ is measured in $(J s m/C)^2$.  Here we have used the convention $x^\mu \equiv (ct,\vec{r}) = (ct,x,y,z)$, $\partial_\mu \equiv (\partial_{ct},\partial_x,\partial_y,\partial_z)$ and $k^\mu \equiv (\omega/c,\vec{k})$, where we are actually working in wave number space rather than momentum space but the two are related by a factor of $\hbar$ as usual (we will often use the word momentum in place of wave number in this article).  As a side note, we remark that their Fourier transforms, defined as $D_{F\mu\nu}(x) = \int d^4k/(2\pi)^4 D_{F\mu\nu}(k) \mbox{exp}(-i k\cdot x)$  and $S_F(x) = \int d^4k/(2\pi)^4 S_F(k) \mbox{exp}(-i k\cdot x)$, are the corresponding Green's functions for the equations
 $\partial^\alpha\partial_\alpha D_{F\mu\nu} (x) = i\left(4\pi k_e\hbar/c\right) \eta_{\mu\nu} \delta^4(x)$ and $\left(i \gamma^\mu \partial_\mu - mc/\hbar\right) S_F(x) = i\delta^4(x)$.  These Green's functions can be used to solve the coupled classical Maxwell-Dirac equations, obtained from the Euler-Lagrange equations for the fields:  
\begin{eqnarray}
\left(i \hbar \gamma^\mu \partial_\mu - m c\right) \psi(x) &=& - e_b \gamma^\mu A_\mu(x) \psi(x)\\
\partial^\nu\partial_\nu A_\mu(x)  &=&   \frac{e_b 4 \pi k_e}{c} \bar{\psi}(x) \gamma_\mu \psi(x) .
\end{eqnarray}

We begin with the Lagrangian of QED, which as a function of the three quantum fields $\psi, \bar{\psi}$ and $A^\nu(x) = (\phi/c,\vec{A})$, takes the form:
\begin{equation}
\mathcal{L}(x) = c\ \bar{\psi}(x) \left(i \hbar \gamma^\mu\partial_\mu - m c\right)\psi(x) -
e_b c\ \bar{\psi}(x) \gamma^\mu A_\mu(x) \psi(x)
-\frac{c^2}{8\pi k_e} A^\mu(x) \left(\partial_\mu\partial_\nu - \eta_{\mu\nu}\partial^2\right) A^\nu(x)
\label{eq:QED Lagrangian}
\end{equation}
where we have defined the photon field to have the usual units of standard electromagnetic textbooks, resulting in the factor of $k_e$ in the free photon field part of the Lagrangian.  Changing the field operators to their wavenumber representation according to the definition of the Fourier transformation $\psi(x) = \int d^4k/(2\pi)^4 \psi(k) \mbox{exp}(-ik\cdot x)$, the total action $\mathcal{S} \equiv \int dt d^3x \mathcal{L} (x) =  c^{-1}\int d^4x \mathcal{L}(x)$ can be written as the sum of three parts.  The first part takes the form
\begin{eqnarray}
\frac{\mathcal{S}_1}{\hbar} 
&=& \frac{1}{c}\int d^4x c\ \bar{\psi}(x) \left(i \gamma^\mu\partial_\mu - \frac{m c}{\hbar}\right)\psi(x)\nonumber\\
&=& \int d^4x \int \frac{d^4k_1}{(2\pi)^4}\bar{\psi}(k_1) e^{ik_1\cdot x} \left(i \gamma^\mu\partial_\mu - \frac{m c}{\hbar}\right) \int \frac{d^4k_2}{(2\pi)^4}\psi(k_2)e^{-ik_2\cdot x}\nonumber\\
&=& \int d^4x \int\frac{d^4k_1}{(2\pi)^4}\bar{\psi}(k_1) e^{i k_1\cdot x} \left(\gamma^\mu k_{2\mu} - \frac{m c}{\hbar}\right) \int \frac{d^4k_2}{(2\pi)^4}\psi(k_2) e^{-i k_2\cdot x}\nonumber\\
&=& \int \frac{d^4k_1d^4k_2}{(2\pi)^8} \bar{\psi}(k_1) \left(\gamma^\mu k_{2\mu} - \frac{m c}{\hbar}\right) \psi(k_2)  (2\pi)^4 \delta^4(k_2-k_1)\nonumber\\
&=& \int \frac{d^4k}{(2\pi)^4} \bar{\psi}(k)\left(\gamma^\mu k_\mu - \frac{m c}{\hbar}\right)\psi(k)
\end{eqnarray}
The inverse of the operator between $\bar{\psi}$ and $\psi$ times $i$ gives us the propagator $i\left(\gamma^\mu k_\mu - m c/\hbar\right)^{-1} \equiv S_F(k)$.  Similarly, for the Maxwell part we obtain
\begin{eqnarray}
\frac{\mathcal{S}_2}{\hbar} 
&=&  -\frac{1}{c}\int d^4x \frac{c^2}{8\pi k_e \hbar} A^\mu(x)\left(\partial_\mu\partial_\nu - \eta_{\mu\nu}\partial^2\right) A^\nu(x)\nonumber\\
&=&  -\int d^4x \frac{c}{8 \pi k_e \hbar} \int\frac{d^4k_1}{(2\pi)^4} A^\mu(k_1) e^{-i k_1\cdot x} \left(\partial_\mu \partial_\nu - \eta_{\mu\nu}\partial^2\right) \int\frac{d^4k_2}{(2\pi)^4}A^\nu(k_2) e^{-i k_2\cdot x}\nonumber\\
&=& \frac{c}{8\pi k_e \hbar} \int\frac{d^4k_1d^4k_2}{(2\pi)^8} A^\mu(k_1) \left(k_{2μ}k_{2ν} - \eta_{\mu\nu}k_2^2\right) A^\nu(k_2) (2\pi)^4 \delta^4(k_1+k_2) \nonumber\\
&=& \int\frac{d^4k}{(2\pi)^4}\frac{1}{2} A^\mu(-k)\frac{c}{4\pi k_e \hbar} \left(k_\mu k_\nu - \eta_{\mu\nu}k^2\right) A^\nu(k) 
\end{eqnarray}
Although the operator between the photon fields can not be directly inverted, after gauge fixing using the Fadeev-Poppov procedure, it can, giving the following propagator (after multiplying by i) $D_{F\mu\nu} (k)= -i(4\pi k_e \hbar /c)\eta_{\mu\nu}/ k^2$, in the Feynman gauge.  
Similarly the interaction part of the action is given by
\begin{eqnarray}
\frac{\mathcal{S}_3}{\hbar}
&=& -\frac{1}{c} \int d^4x \bar{\psi}(x) \frac{c e_b}{\hbar} \gamma^\mu A_\mu(x) \psi(x)\nonumber\\ &=& -\int d^4x \int\frac{d^4k_1}{(2\pi)^4} \bar{\psi}(k_1) e^{i k_1\cdot x} \frac{e_b}{\hbar} \gamma^\mu \int\frac{d^4k_2}{(2\pi)^4} A_\mu(x) e^{-i k_2\cdot x} \int\frac{d^4k_3}{(2\pi)^4} \psi(k_3) e^{-i k_3\cdot x}\nonumber\\ 
&=& \int\frac{d^4k_1d^4k_2d^4k_3}{(2\pi)^{12}} \bar{\psi}(k_1) \frac{-e_b}{\hbar}\gamma^\mu A_\mu(k_2) \psi(k_3) (2\pi)^4 \delta^4(k_3+k_2-k_1)
\end{eqnarray}
We extract the vertex operator after functionally differentiating $\mathcal{S}_3/\hbar$ with respect to the three fields and multiplying by $i$ giving us $V^\mu= -i e_b \gamma^\mu/\hbar$.
\begin{equation}
\frac{\delta \mathcal{S}_3}{\delta\psi(k_3)\delta A_\mu(k_2)\delta\bar{\psi}(k_1)} = -\frac{e_b}{\hbar}\gamma^\mu (2\pi)^4\delta^4(k_3+k_2-k_1)
\end{equation}

\section{\label{app:quantum correction to potential}The Order-$\mathbf{\alpha}$ Quantum Correction to the Electric Potential}
\begin{figure}
\begin{center}
\begin{minipage}{1.25in}
\includegraphics[scale=0.5]{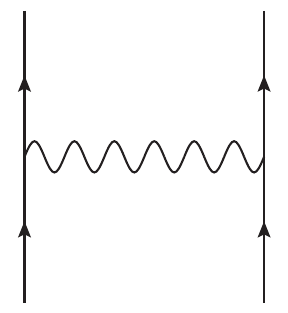}\\\textbf{(a)}
\end{minipage}
\begin{minipage}{1.25in}
\includegraphics[scale=0.5]{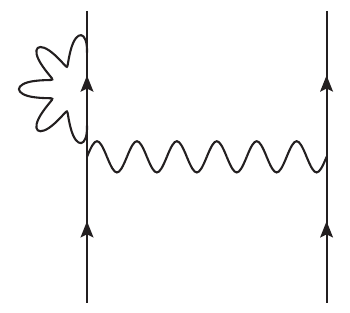}\\(b)
\end{minipage}
\begin{minipage}{1.25in}
\includegraphics[scale=0.5]{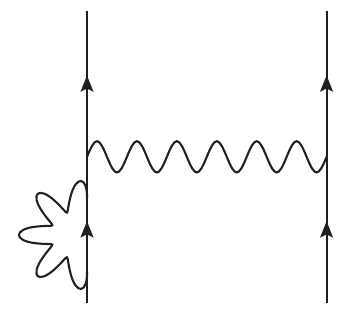}\\(c)
\end{minipage}
\begin{minipage}{1.25in}
\includegraphics[scale=0.5]{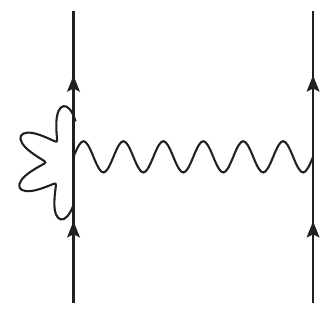}\\(d)
\end{minipage}
\begin{minipage}{1.25in}
\includegraphics[scale=0.5]{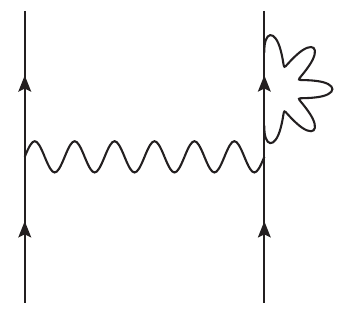}\\(e)
\end{minipage}\\
\begin{minipage}{1.25in}
\includegraphics[scale=0.5]{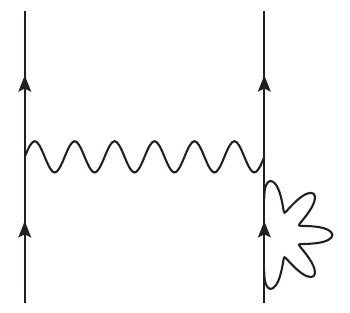}\\(f)
\end{minipage}
\begin{minipage}{1.25in}
\includegraphics[scale=0.5]{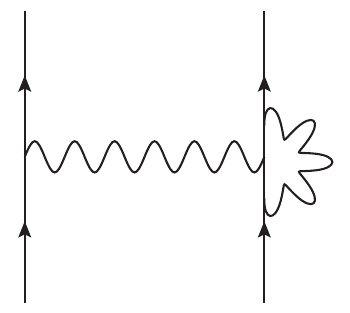}\\(g)
\end{minipage}
\begin{minipage}{1.25in}
\includegraphics[scale=0.5]{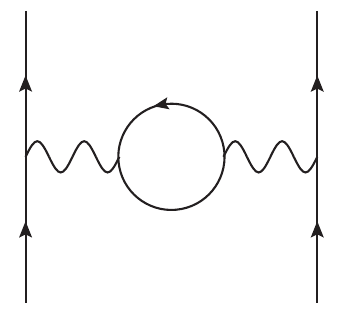}\\\textbf{(h)}
\end{minipage}
\begin{minipage}{1.25in}
\includegraphics[scale=0.5]{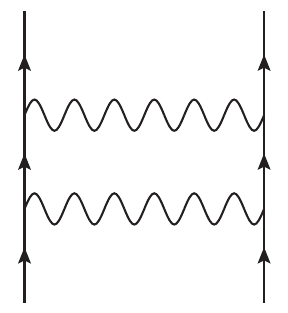}\\(i)
\end{minipage}
\begin{minipage}{1.25in}
\includegraphics[scale=0.5]{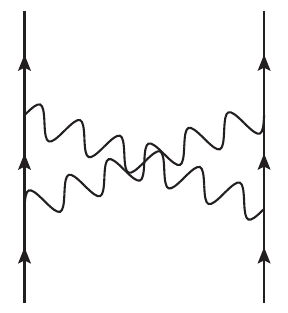}\\(j)
\end{minipage}
\end{center}
\caption{\label{fig:complete diagrams}Leading order diagram (a) and one-loop diagrams (b-j) that contribute to the electric potential between charge carriers in the two plates.  In the long-range limit (low energy transfer limit) we consider, the potential is dominated by diagrams (a) and (h).  The wavy lines represent photons and the straight lines represent electrons.}
\end{figure}

In this section, we would like to calculate the leading-order correction to the electric potential of a charged particle.  This is a standard textbook calculation \cite{Peskin:1995ev} and we only outline the derivation here, focusing on aspects that will be important in our 1d calculation.  In Figure \ref{fig:complete diagrams}, we list all nine one-loop diagrams that potentially contribute at next-to-leading order to the interaction between the charge carriers.  For reference, we also include the leading-order (tree-level) diagram as Figure \ref{fig:complete diagrams}(a).  In order to understand the contributions from these diagrams, we begin by noting that we are interested in the long-range behavior of the electric potential (distances on the order of or greater than the Compton wavelength of the electron).  In Fourier-transformed space, this corresponds with very low energy being transferred between the charge carriers in the parallel plate.  In this limit, the digrams in Figures \ref{fig:complete diagrams}(b)-(d) exactly cancel (there are also bremstrahlung diagrams which are important for this cancellation which will not be important in our calculation).  This exact cancellation is due to the Ward identity which is a consequence of the QED gauge symmetry.  The same cancellation occurs at long range for the diagrams in Figures \ref{fig:complete diagrams}(e)-(g).  (These diagrams are only important for short-range interactions.)  The diagrams in Figures \ref{fig:complete diagrams}(i)-(j) are finite, but are suppressed by the mass of the electron.  As a result, these diagrams are only important at short range (smaller than the Compton wavelength) when the energy exchanged between the charge carriers is greater than $m c^2$.  This leaves us with only the diagram in Figure \ref{fig:complete diagrams}(h), which does contribute at long range as well as at short range.

\begin{figure}
\begin{center}
\begin{minipage}{1.5in}
\begin{center}
\includegraphics[scale=0.5]{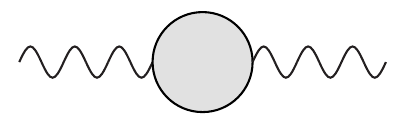}
\end{center}
\end{minipage}
\begin{minipage}{0.4in}
\begin{center}
=
\end{center}
\end{minipage}
\begin{minipage}{1.25in}
\includegraphics[scale=0.5]{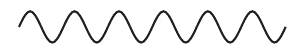}
\end{minipage}
\begin{minipage}{0.2in}
+
\end{minipage}
\begin{minipage}{1.5in}
\includegraphics[scale=0.5]{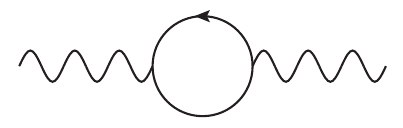}
\end{minipage}
\end{center}
\caption{\label{fig:modified photon prop}Modified photon propagator at one loop.  It is the modified photon propagator that will contribute to the long-range electric potential.  The wavy line with the shaded circle on the left represents the combined effects of the tree-level photon propagator shown as the first term on the right and the one-loop correction to the photon propagator shown as the second term on the right.}
\end{figure}

Since the quantum correction to the long-range potential is dominated by the diagram in Figure~\ref{fig:complete diagrams}(h), we now focus on just this diagram and the leading order diagram in Figure~\ref{fig:complete diagrams}(a).  We see that they can both be combined into one diagram with a modified photon propagator as shown in Figure \ref{fig:modified photon prop}.
We can construct the modified photon propagator from these diagrams by writing
\begin{equation}
D_{F}^{\prime\mu\nu}(k) =  D_{F}^{\mu\nu}(k) + D_{F}^{\mu\sigma}(k)\left[i\Pi_{\sigma\lambda}(k)\right]D_F^{\lambda\nu}(k)
\end{equation}
where $D_F^{\prime\mu\nu}(k)$ represents the modified photon propagator, $D_F^{\mu\nu}(k)$ is the leading-order photon propagator from Appendix \ref{app:Feynman Rules} and $i\Pi_{\sigma\lambda}(k)$, called the polarization tensor, is the result of the loop, which we will calculate now.  We take the loop momentum $p$ to be oriented in the direction of the arrow on the loop while we choose the propagator momentum $k$ to travel along the lower half of the loop to the other side.  The polarization tensor is then determined, following standard Feynman rules, by following the fermion loop around in the opposite direction of the arrows writing down factors as we go.  We obtain
\begin{equation}
i\Pi_{\sigma\lambda}(k) = -\int\frac{d^4p}{(2\pi)^4}\mbox{Tr}\left[V_\sigma S_F(p) V_\lambda S_F(p+k)\right]
\label{eq:polarization tensor start}
\end{equation}
where the minus sign comes from having a loop of fermions and the trace is over Dirac gamma matrices.

Unfortunately the integral over p does not converge.  However, it turns out that the singularity can be absorbed into the definition of the unobservable bare coupling constant $e_b$.  In order to do this, the integral must first be regularized, typically by dimensional regularization, which preserves the Ward identities.  This is done by replacing the 4-dimensional integral by the limit of a $(4-\epsilon)$ dimensional integral as $\epsilon\to0$.  However, the $\epsilon\to0$ limit is taken after absorbing the potentially infinite terms into the bare coupling $e_b$.  With this, after taking the trace, combining the denominators and making a change of integration variables, the polarization tensor can be put in the form
\begin{equation}
\Pi_{\sigma\lambda}(k) = \frac{4i e_b^2}{\hbar^2} \int_0^1d\beta\int\frac{d^{4-\epsilon}p}{(2\pi)^{4-\epsilon}}
\frac{2p_\sigma p_\lambda - p^2\eta_{\sigma\lambda} - 2\beta(1-\beta)k_\sigma k_\lambda+\beta(1-\beta)k^2\eta_{\sigma\lambda}+\left(m^2c^2/\hbar^2\right)\eta_{\sigma\lambda}}{\left(p^2+\beta(1-\beta)k^2 - m^2c^2/\hbar^2\right)^2}
\label{eq:unintegrated polarization tensor}
\end{equation}
The analytical form of these integrals can be looked up in standard references such as \cite{Peskin:1995ev} giving us
\begin{equation}
\Pi_{\sigma\lambda}(k) = - \frac{8e_b^2}{(4\pi\hbar)^2}\left(\eta_{\sigma\lambda} k^2-k_\sigma k_\lambda\right) 
\int_0^1 d\beta \beta(1-\beta) \left[\frac{2}{\epsilon} -\gamma + \mbox{ln}\left(\frac{4\pi}{\Delta}\right) + \mathcal{O}(\epsilon)\right]
\end{equation}
where $\Delta=m^2c^2/\hbar^2-\beta(1-\beta)k^2$.  As a test of our expression, we immediately see that $k^\sigma\Pi_{\sigma\lambda}(k)=0$, as required by the Ward identities.  To simplify our notation we define $P(k^2)$, where we have factored out the bare coupling,  $\Pi_{\sigma\lambda}(k)\equiv e_b/(4\pi\hbar^2)\left(\eta_{\sigma\lambda}k^2 -k_\sigma k_\lambda\right)P(k^2)$.  As a result, the modified propagator takes the form:
\begin{equation}
D'_{F\mu\nu}(k) = -i\frac{4\pi k_e \hbar}{c}\frac{\eta_{\mu\nu}}{k^2}
-i \frac{e_b^2}{4\pi\hbar^2} \left(\eta_{\mu\nu}k^2-k_\mu k_\nu\right) P(k^2) \left(\frac{4\pi k_e\hbar k^2}{c}\right)^2
\end{equation}
In our parallel plate capacitor, the ends of this propagator will be connected to the charge carriers in the two plates (the straight lines on the two sides in Figure \ref{fig:complete diagrams}) in terms such as $\bar{u}(p+k)\gamma^\mu u(p) D'_{F\mu\nu}(k)$.  When the $k_\mu k_\nu$ term of the modified photon propagator combines with the gamma matrix, we will get $\bar{u}(p+k) [\gamma^\mu(p_\mu+k_\mu)- \gamma^\mu p_\mu] u(p)$.  However, since $\bar{u}(p+k) \gamma^\mu(p_\mu+k_\mu) = \bar{u}(p+k)m$ and $\gamma^\mu p_\mu u(p) =m u(p)$ (these are the Euler-Lagrange equations for the Dirac spinor in momentum space), we see that the contribution from $k_\mu k_\nu$ vanishes.  This is part of a larger Ward identity that states that at the one-loop level we are working at, the $k_\mu k_\nu$ piece does not contribute to any physical observables.  Therefore, we can drop it and we find
\begin{equation}
D'_{F\mu\nu}(k) = D_{F\mu\nu}(k) \left[1+ \frac{e_b^2 k_e}{\hbar c} P(k^2)\right]
\end{equation}
where we remind the reader that $D_{F\mu\nu}(k) = -i(4\pi k_e\hbar/c) \eta_{\mu\nu}/k^2$.  In this form, we see that the result of this quantum correction is nothing but a momentum dependent correction factor due to quantum one-loop effects.
	In order to determine the (so far) unknown value of the bare coupling parameter $e_b$, we need a specific experimentally measureable case as a reference.  We will convert the modified Feynman propagator to a potential associated with a general bare source at long range.  We will introduce this bare source current as $j_b^\nu(k)= e_b j_t^\nu(k)$ where we have factored out the bare charge and note that we assume the remaining $j_t^\nu(k)$ is spatially localized.  For example, a point charge at rest would be given by $j_t^\nu(k)=c 2\pi\delta(k_0)\eta^{\nu0}$.  In order to do this, we will use Green's function to give the quantum corrected potential associated with the bare charge as
\begin{equation}
A'_\mu(k) = -i\frac{1}{\hbar c} D'_{F\mu\nu}(k) e_b j_t^\nu(k)
\end{equation}
We then obtain
\begin{equation}
  A'_\mu(k)  = -i \frac{1}{\hbar c} D_{F\mu\nu}(k) \left[1+ \frac{e_b^2 k_e}{\hbar c} P(k^2)\right] e_b j_t^\nu(k)
\label{eq:app B:A' eb}
\end{equation}
The corresponding (experimentally verified at long range) classical potential for a physical charge given by $j^\nu(k)= e j_t^\nu(k)$, where it is important that $j_t^\nu(k)$ is the same as above and $e= 1.6\times10^{-19}$C is the measured value of the positron charge, is
\begin{equation}
A_\mu(k) = -i\frac{1}{\hbar c} D_{F\mu\nu}(k) e j_t^\nu(k)
\label{eq:app B:A classical}
\end{equation}
In order for the long-range behavior of the potential to be the same in these two theories, we must equate the limit as $k\to0$ of Eqs. (\ref{eq:app B:A' eb}) and (\ref{eq:app B:A classical})
\begin{equation}
\lim_{k\to0} A'_\mu(k)  =   \lim_{k\to0} A_\mu(k)
\end{equation}
and therefore, we find
\begin{equation}
e = e_b\left[1+ \frac{e_b^2 k_e}{\hbar c} P(0)\right]
\end{equation}
If we multiply both sides by $\sqrt{k_e/(\hbar c)}$, we have
\begin{equation}
\alpha^{1/2} = \alpha_b^{1/2}\left[1+ \alpha_b P(0)\right]
\end{equation}
where we have defined the unitless $\alpha\equiv e^2k_e/(\hbar c)$ and $\alpha_b\equiv e_b^2k_e/(\hbar c)$.  Note that $\alpha\simeq 1/137$ is the fine structure constant of 3d QED.  
Solving this equation order by order in $\alpha^{1/2}$, we find up to order $\alpha^{3/2}$ (and removing the factor $k_e/(\hbar c)$ on both sides)
\begin{equation}
e_b = e\left[1-\alpha P(0)\right] 
\end{equation}
 We note that formally $P(0)$ is infinite and therefore $e_b$ is also infinite.
	We can now insert this expression back into Eq. (\ref{eq:app B:A' eb}) to obtain the renormalized potential for other values of $k$ 
\begin{equation}
A'_\mu(k)  =  -i\frac{1}{\hbar c} D_{F\mu\nu}(k) \left(1+ \alpha\left[P(k^2)-P(0)\right]\right) j^\nu(k)
\end{equation}
Here, we see that physically observable quantities do not depend on $P(k^2)$ alone, but on the difference $P(k^2)-P(0)$, 
\begin{equation}
P(k^2) - P(0) = \frac{2}{\pi} \int_0^1d\beta\beta(1-\beta) \mbox{ln}\left[ 1 -\beta(1-\beta)k^2\lambda_C^2\right]
\end{equation}
where $\lambda_C \equiv \hbar/(m c)$  ($=3.85\times10^{-13}$m) is the Compton wave length.  We note that this is finite in the limit $\epsilon\to0$, which is technically due to the cancellation of the $2/\epsilon$ terms.


Now that we have the modified potential in momentum space, we would like to Fourier transform it back to position space.  In particular, we would like to consider the potential due to a point charge at rest $j^\nu(x)=e c \eta^{\nu 0}\delta^3(\vec{r})$.  Plugging this in, we obtain
\begin{eqnarray}
A'_\mu(k)
&=& -i\frac{1}{\hbar c}D'_{F\mu\nu}(k)\int d^4x e^{i k\cdot x} j^\nu(x)\nonumber\\
&=& -i\frac{e}{\hbar}D'_{F\mu 0}(k) (2\pi)\delta(k^0)
\end{eqnarray}
Fourier transforming this result gives
\begin{eqnarray}
\phi(\vec{r}) 
&=& -i\frac{e c}{\hbar}\int\frac{d^4k}{(2\pi)^4}D'_{F00}(k)(2\pi)\delta(k^0)e^{-i k\cdot x}\nonumber\\
&=& -i\frac{e c}{\hbar}\int\frac{d^3k}{(2\pi)^3}D'_{F00}(\vec{k})e^{-i\vec{k}\cdot\vec{r}}\nonumber\\
&=& 4\pi k_e e \int\frac{d^3k}{(2\pi)^3}\frac{1}{\vec{k}^2}\left(1+\frac{2\alpha}{\pi}\int_0^1d\beta\beta(1-\beta)\mbox{ln}\left[1+\beta(1-\beta)\vec{k}^2\lambda_C^2\right]\right)e^{-i\vec{k}\cdot \vec{r}}
\end{eqnarray}
The integral of the first term is $1/(4\pi r)$ giving the standard classical Coulomb potential $k_e e/r$.  In order to integrate the second term, we convert to spherical coordinates and integrate over the angles to obtain
\begin{equation}
\delta \phi(r) = -i\frac{k_e e \alpha}{\pi^2r} \int_{-\infty}^\infty \frac{dq}{q}\left(e^{i q r}-e^{-i q r}\right)\  \int_0^1d\beta\beta(1-\beta) \mbox{ln}\left[1+\beta(1-\beta)q^2\lambda_C^2\right]
\end{equation}
where $q=|\vec{k}|$ and we have used the fact that the integrand is even in $q$ to write the integral $\int_0^\infty$ as $\frac{1}{2}\int_{-\infty}^\infty$.  In order to complete the integral over $q$, we analytically continue the integrand to the complex $q$ plane.  We note that there are no poles in the integrand (both the numerator and denominator go to zero as $q\to0$), but there is a branch cut beginning at $q=\pm i/[\lambda_C\sqrt{\beta(1-\beta)}]$ and continuing up to $q\to \pm i\infty$.  We will split this integral into the two pieces
\begin{equation}
I_\pm = \int_{-\infty}^\infty \frac{dq}{q}e^{\pm i q r}\int_0^1d\beta\beta(1-\beta) \mbox{ln}\left[1+\beta(1-\beta)q^2\lambda_C^2\right]
\end{equation}
For each, we will add to this integral the half circle at complex infinity in the upper half plane for $I_+$ and in the lower half plane for $I_-$.  This contributes nothing to the integral because of the suppression in the exp$(\pm i q r)$ term.  However, when we reach the branch cut, we will need to integrate down the branch cut towards the origin until we reach the end and then back up the branch cut on the other side.  Since these contour integrals will not enclose any poles, they will be zero showing that our original integrals $I_\pm$ are equal to minus the integrals around the branch cuts.  Since the real part of the integrand is the same on the two sides of the branch cut, they will cancel.  The imaginary part of the logarithm, on the other hand, will differ by $2\pi$ between the two sides of the branch cut, giving us
\begin{equation}
I_\pm = \pm2 i\pi \int_{2/\lambda_C}^{\infty}\frac{dq}{q}e^{- q r}\int_{\frac{1}{2}\left[1-\sqrt{1-4/(\lambda_Cq)^2}\right]}^{\frac{1}{2}\left[1+\sqrt{1-4/(\lambda_Cq)^2}\right]} d\beta\beta(1-\beta) 
\end{equation}
where we have only integrated over the imaginary part of the logarithm.  Performing the $\beta$ integral and plugging back into $\delta \phi(r)$ gives us
\begin{equation}
\delta\phi(r) =\frac{k_e e }{r}\frac{\alpha}{3\pi}\int_{2/\lambda_C}^{\infty}\frac{dq}{q}e^{- q r}\left[2+\left(\frac{2}{\lambda_Cq}\right)^2\right]\sqrt{1-\frac{4}{\lambda_C^2q^2}}
\end{equation}
Finally, making the change of variables $q=2\tau/\lambda_C$ and including the leading-order expression again gives us
\begin{equation}
\phi(r) =\frac{k_e e}{r}\left[1+ \frac{\alpha}{3\pi } \int_1^\infty d\tau f(\tau) e^{-2\tau r/\lambda_C}\right]
\label{eq:Uehling potential}
\end{equation}
where we have defined
\begin{equation}
f(\tau) \equiv \left(\frac{2}{\tau^2}+\frac{1}{\tau^4}\right)\sqrt{\tau^2-1}
\end{equation}
This agrees with the well-known form of the Uehling potential \cite{Uehling:1935uj,Wichmann:1956zz,Fullerton:1976fu,Soff:1988zz,Neghabian:1983,Indelicato:2014mra}.  We see that the second term in the Uehling potential is higher order in the coupling constant ($e^3$) than the the Coulomb term (order $e$), so that this corresponds with a perturbation in the electric charge, as expected.  We have derived the potential for a point charge but the generalization to a charge density is clear
\begin{equation}
\phi(r)  = k_e \int d^3r' \frac{\rho(\vec{r}^{\ \prime})}{|\vec{r}-\vec{r}^{\ \prime}|} 
\left[1 + \frac{\alpha}{3\pi}\int_1^\infty d\tau f(\tau)e^{-2\tau |\vec{r}-\vec{r}^{\ \prime}|/\lambda_C}\right]
\end{equation}

\section{\label{app:1d QED}One-Dimensional QED}
In this section, we begin by defining what we mean by a 1d theory.  We do \textit{not} mean that we start from scratch with one dimension of space (and one dimension of time) and construct field theory.  What we mean in this article is that we begin with a fundamental three-dimensional theory (plus time) and reduce the theory by removing the dependence on two of the dimensions.  Without loss of generality, let's suppose that the dimensions we will drop are the $x_1$ and $x_2$ dimensions and we will now use the notation $x=(ct,z)$.  We reduce the Lagrangian by dropping all derivatives with respect to $x_1$ and $x_2$, dropping all dependence on $x_1$ and $x_2$ in the fields, dropping the fields $A^1(x)$ and $A^2(x)$ and removing dependence on the Dirac gamma matrices $\gamma^1$ and $\gamma^2$.  We are also guided in the way we formulate this theory by keeping the same value and units for the speed of light $c$, the electron mass $m$ and Coulomb's constant $k_e$.  We can accomplish this by defining our 1d fields with an absorbed factor of $L$ where $L^2=\int dx_1 dx_2$ is the infinite area along the $x_1$ and $x_2$ directions.  So, $\psi^{(1d)}(ct,z)=L\psi(ct,z)$ where $\psi(ct,z)$ is the 3d field with the dependence on $x_1$ and $x_2$ dropped.  We still define $\bar{\psi}^{(1d)}=\psi^{(1d)\dagger}\gamma^0$ as in the 3d theory. We also define $A^{(1d)\mu}(ct,z)=L A^\mu(ct,z)$ where we have dropped dependence on $x_1$ and $x_2$ and multiplied by $L$ and the index $\mu$ only takes the values $0$ and $3$.  We will assume that all Lorentz indices will only take the values $0$ and $3$ for the rest of this section.  We then find that our 1d Lagrangian is given by multiplying the orginal 3d Lagrangian (where all derivatives or momenta along the $x_1$ and $x_2$ directions were ommited) by a factor of $L^2$, leading to
\begin{equation}
\mathcal{L}^{(1d)}(x) = c\bar{\psi}^{(1d)}\left(i\hbar\gamma^\mu\partial_\mu-m c\right)\psi^{(1d)}
-\kappa_b c\bar{\psi}^{(1d)}\gamma^\mu A^{(1d)}_\mu\psi^{(1d)}
-\frac{c^2}{8\pi k_e}A^{(1d)\mu}\left(\partial_\mu\partial_\nu-\eta_{\mu\nu}\partial^2\right)A^{(1d)\nu}
\label{eq:appC:L1d}
\end{equation}
In addition to the redefinition of the fields, we have also introduced a new coupling constant $\kappa_b=e_b/L$.  All other constants remain the same.  The action is given by the integral $\mathcal{S}^{(1d)}=\int dt dz \mathcal{L}^{(1d)} = c^{-1}\int d^2x\mathcal{L}^{(1d)}$ and the Feynman rules are obtained by the same procedure as in Appendix \ref{app:Feynman Rules} giving
\begin{eqnarray}
V^{(1d)}_\mu &=& -i \frac{\kappa_b}{\hbar}\gamma_\mu\\
S^{(1d)}_F(k) &=& \frac{i}{\gamma^\mu k_\mu-m c/\hbar}\\
D^{(1d)}_{F\mu\nu} &=& -i \frac{4\pi k_e\hbar}{c}\frac{\eta_{\mu\nu}}{k^2}
\end{eqnarray}
where the main differences (beyond a restriction of the Lorentz indices to $0$ and $3$) are the replacement of $e_b$ with $\kappa_b$, the implicit momentum conserving delta function is now only over two dimensions $(2\pi)^4\delta^4(\sum k) \to (2\pi)^2\delta^2(\sum k)$ and we use the property that any $k_\mu k_\nu$ contribution to the propagator will not contribute to any physical observables at the order of perturbation theory we are working.  As expected, the propagators are the Green's functions that satisfy $\partial^\alpha\partial_\alpha D_{F\mu\nu} (x) = i\left(4\pi k_e\hbar/c\right) \eta_{\mu\nu} \delta^4(x)$ and $\left(i \gamma^\mu \partial_\mu - mc/\hbar\right) S_F(x) = i\delta^4(x)$.  We note that the units for the vertex, given by $\kappa_b/\hbar$, are $C/J$, the units for the electromagnetic field $A^{(1d)\mu}$ are $J s/C$ and the units for the fermion field are $m^{-1/2}$.

Any 3d system whose charge distribution depends only on the $z$-direction must naturally be infinitely extended along the $x_1$- as well as the $x_2$-direction.  As a result, all total forces $F$ and energies $V$ are infinite, but one can still compute the corresponding (finite) two-dimensional densities such as $F/L^2$ or $V/L^2$, where we are denoting the (in principle infinite) area by the quantity $L^2$.  For example, as we have seen in the main text, the (finite) charge density of a plate can be characterized by $q$, measured in units of $C/m^2$. 
  
In addition to the new quantities defined with superscripts $(1d)$, we also need to articulate translation rules in order to compare the 1d and 3d observables quantitatively.  The total 1d charge [defined as $\int dz \rho^{(1d)}(z)$] needs to be multiplied by $L$ and the electric field [defined as $(-\partial_z\phi^{(1d)})/q^{(1d)}$] needs to be multiplied by $L^2$ to predict the corresponding quantitities in the 3 d world.  With these rules, forces will have the usual units of $N$.

We would now like to calculate the correction to the Coulomb potential using our 1d theory in order to compare with the predictions of the full 3d theory.  We assume that we can still neglect all the diagrams of Fig. \ref{fig:complete diagrams} except Figs. \ref{fig:complete diagrams}(a) and (h).  Combining these two diagrams as in Appendix \ref{app:quantum correction to potential}, we calculate a modified propagator for the photon
\begin{equation}
D_{F}^{(1d)\prime\mu\nu}(k) =  D_{F}^{(1d)\mu\nu}(k) + D_{F}^{(1d)\mu\sigma}(k)\left[i\Pi^{(1d)}_{\sigma\lambda}(k)\right]D_F^{(1d)\lambda\nu}(k)
\end{equation}
Using our 1d Feynman rules, we obtain
\begin{equation}
i \Pi^{(1d)}_{\sigma\lambda}(k) = -\int\frac{d^2p}{(2\pi)^2}\mbox{Tr}\left[V^{(1d)}_\sigma S^{(1d)}_F(p) V^{(1d)}_\lambda S^{(1d)}_F(p+k)\right]\\
\end{equation}
This integral is formally logarithmically divergent and we must regularize it to proceed.  We will again use dimensional regularization as it preserves the Ward identities.  We change the integration from $d^2p$ to $d^{2-\epsilon}p$, trace the gamma matrices, combine the propagator denominators and change variables to obtain
\begin{equation}
\Pi^{(1d)}_{\sigma\lambda}(k) = \frac{4i\kappa_b^2}{\hbar^2}\int_0^1d\beta\int \frac{d^{2-\epsilon}p}{(2\pi)^{2-\epsilon}}
\frac{2p_\sigma p_\lambda-p^2\eta_{\sigma\lambda}-2\beta(1-\beta)k_\sigma k_\lambda+\beta(1-\beta)k^2\eta_{\sigma\lambda}+\eta_{\sigma\lambda}m^2c^2/\hbar^2}{\left(p^2+\beta(1-\beta)k^2-m^2c^2/\hbar^2\right)^2}
\end{equation}
which is very similar to Eq.~(\ref{eq:unintegrated polarization tensor}) although we will find that the result of this integration will be very different.  Looking these integrals up in a standard reference \cite{Peskin:1995ev}, we obtain
\begin{equation}
\Pi^{(1d)}_{\sigma\lambda}(k) = 
-\frac{2\kappa_b^2}{\pi\hbar^2}\left(k^2\eta_{\sigma\lambda}-k_\sigma k_\lambda\right)
\int_0^1d\beta \frac{\beta(1-\beta)}{m^2c^2/\hbar^2-\beta(1-\beta)k^2}
\end{equation}
In this case, because of the smaller dimension, the result is independent of $\epsilon$.  All formally divergent terms (in the limit of $\epsilon\to0$) exactly cancel.  This is because we used a regularization scheme that preserved the Ward identities and the Ward identities demanded that the result be proportional to $k^2\eta_{\sigma\lambda}-k_\sigma k_\lambda$.  After factoring this out of the regularized integral, it was reduced from a logarithmically divergent integral $\int d^2p/p^2$ to a convergent integral $\int d^2p/p^4$.  (In the 3d theory, on the other hand, the integral was reduced from a quadratically divergent integral $\int d^4p/p^2$ to a logarithmically divergent integral $\int d^4p/p^4$ so that a $1/\epsilon$ pole remained after the regularization.)  In order to simplify our notation, we again introduce a function we call $P^{(1d)}(k^2)$ where we have factored out the bare coupling constant  $\Pi^{(1d)}_{\sigma\lambda}(k)\equiv\kappa_b^2/(4\pi\hbar^2)\left(\eta_{\sigma\lambda} k^2-k_\sigma k_\lambda\right)P^{(1d)}(k^2)$.  We also note that the $k_\sigma k_\lambda$ term will not contribute to any physical observables at this order in the coupling constant and so drop it as we did in Appendix \ref{app:quantum correction to potential}.  With these two simplifications, we can write the next-to-leading-order photon propagator as
\begin{equation}
D^{(1d)\prime}_{F\mu\nu}(k) = D^{(1d)}_{F\mu\nu}(k)\left[1+\frac{\kappa_b^2k_e}{\hbar c}P^{(1d)}(k^2)\right]
\end{equation}

We next need to determine the value of the bare coupling $\kappa_b$ by relating it to a known coupling.  We do this by following the same procedure as in the previous section.  We convert the modified Feynman propagator to a potential associated with a general bare source at long range.  We again introduce a bare source current as $j_b^{(1d)\nu}(k)=\kappa_b j_t^{(1d)\nu}(k)$ where we have factored out the bare charge $\kappa_b$ and note that we assume the remaining $j_t^\nu(k)$ is spatially localized in the 1 dimension.  For example, a ``point charge'' at rest would be given by $j_t^\nu(x) = c\eta^{\nu0}\delta(z)$.  As in the previous section, we use Green's function to give the quantum corrected potential associated with the bare charge as
\begin{equation}
A^{(1d)'}_\mu(k) = -i\frac{1}{\hbar c}D^{(1d)'}_{F\mu\nu}(k)\kappa_b j_t^{(1d)\nu}(k)
\end{equation}
from which we obtain
\begin{equation}
A^{(1d)'}_\mu(k) = -i\frac{1}{\hbar c}D^{(1d)}_{F\mu\nu}(k) \left[1+\frac{\kappa_b^2k_e}{\hbar c}P^{(1d)}(k^2)\right] \kappa_b j_t^{(1d)\nu}(k)
\label{eq:AppC:A' expanded}
\end{equation}
The corresponding classical potential for a physical charge given by $j^{(1d)\nu}(k) = e^{(1d)}j_t^{(1d)\nu}(k)$, where $e^{(1d)}$ is not yet known in contrast to the 3d charge, is
\begin{equation}
A^{(1d)}_\mu(k) = -i\frac{1}{\hbar c}D^{(1d)}_{F\mu\nu}(k) \kappa_b j_t^{(1d)\nu}(k)
\label{eq:AppC:A classical}
\end{equation}
In order for the long-range behavior of the potential to be the same in these two theories, we must equate the limit as $k\to0$ of Eqs. (\ref{eq:AppC:A' expanded}) and (\ref{eq:AppC:A classical})
\begin{equation}
\lim_{k\to0}A^{(1d)'}_\mu(k) = \lim_{k\to0} A^{(1d)}_\mu(k)
\end{equation}
As in the previous section, this results in the relationship
\begin{equation}
e^{(1d)} = \kappa_b\left[1 + \kappa_b^2 \frac{k_e}{c\hbar}P^{(1d)}(0)\right]
\end{equation}
Solving this order by order in $e^{(1d)}$ to second order, as we did in the previous section, we obtain
\begin{equation}
\kappa_b = e^{(1d)}\left[1-\alpha^{(1d)}P^{(1d)}(0)\right]
\end{equation}
where we have defined $\alpha^{(1d)}\equiv k_e [e^{(1d)}]^2/(\hbar c)$.  We note that, in contrast to the 3d theory, here, $\kappa_b$ is finite since $P^{(1d)}(0)$ is finite.  Plugging this expression for $\kappa_b$ back into the Green's function, we obtain
\begin{equation}
A^{(1d)\prime}_\mu(k) = -i\frac{1}{\hbar c}D^{(1d)}_{F\mu\nu}(k)\left(1+\alpha^{(1d)}\left[P^{(1d)}(k^2)-P^{(1d)}(0)\right]\right)j^{(1d)\nu}(k)
\end{equation}
where
\begin{equation}
P^{(1d)}(k^2)-P^{(1d)}(0) = -8\lambda_C^2\int_0^1d\beta\frac{\beta^2(1-\beta)^2k^2\lambda_C^2}{1-\beta(1-\beta)k^2\lambda_C^2}
\end{equation}
where we have again used the Compton wavelength $\lambda_C=\hbar/(m c)$.

Now that we have the complete 1d Green's function, we would like to apply it to the specific case of a 1d ``point charge'', namely $j^{(1d)\nu}(x)=e^{(1d)}c\eta^{\nu0}\delta(z)$.  Plugging this in, we have
\begin{eqnarray}
A^{(1d)\prime}_\mu(k) 
&=& -i\frac{e^{(1d)}}{\hbar}D^{(1d)\prime}_{F\mu 0}(k)(2\pi)\delta(k^0)
\end{eqnarray}
Fourier transforming this result gives
\begin{eqnarray}
\phi^{(1d)}(z) 
&=& 4\pi k_e e^{(1d)}\int\frac{d k_z}{2\pi}\frac{1}{k_z^2}\left[1+8\alpha^{(1d)}\lambda_C^2
\int_0^1d\beta\frac{\beta^2(1-\beta)^2k_z^2\lambda_C^2}{1+\beta(1-\beta)k_z^2\lambda_C^2}\right]
e^{-i k_z z}
\end{eqnarray}
The first term gives $-2\pi k_e e^{(1d)}|z|$ as expected from the classical equations.  The second integral can be performed by analytically continuing $k_z$ to the complex plane and adding the half circle at complex infinity in the upper half plane for $z<0$ and in the lower half plane for $z>0$ which contributes nothing due to the supression of the exp$(-i k_z z)$ term.  We note that there are simple poles in the integrand at the values $k_z = \pm i/\left[\lambda_C\sqrt{\beta(1-\beta)}\right]$.  The integral gives $2\pi i$ times the residue enclosed in the contour giving us
\begin{equation}
\delta\phi^{(1d)}(z) = 16\pi k_e e^{(1d)}\alpha^{(1d)}\lambda_C^3\int_0^1 d\beta 
\left[\beta(1-\beta)\right]^{3/2}e^{-|z|/[\lambda_C\sqrt{\beta(1-\beta)}]}
\end{equation}
Finally, we note that the integrand is symmetric between the two halves $\beta=0$ to $1/2$ and $\beta=1/2$ to $1$ so we replace the $\int_0^1 d\beta$ integral with $2\int_{1/2}^1d\beta$.  We also make a change of variables $2\tau=1/\sqrt{\beta(1-\beta)}$ to obtain
\begin{equation}
\phi^{(1d)}(z) = 2\pi k_e e^{(1d)} \left[-|z| +
\alpha^{(1d)}\lambda_C^3\int_1^\infty d\tau f^{(1d)}(\tau)e^{-2\tau|z|/\lambda_C}\right] 
\label{eq:appC:phi1d}
\end{equation}
where
\begin{equation}
f^{(1d)}(\tau) \equiv \frac{1}{\tau^5\sqrt{\tau^2-1}}
\end{equation}
The generalization of this to a charge distribution is then
\begin{equation}
\phi^{(1d)}(z) = 2\pi k_e \int dz'\rho(z') \left[-|z-z'| +
\alpha^{(1d)}\lambda_C^3\int_1^\infty d\tau f^{(1d)}(\tau)e^{-2\tau|z-z'|/\lambda_C}\right] 
\label{eq:appC:phi1d2}
\end{equation}


\end{document}